\documentclass[aps,superscriptaddress,amsmath,amssymb,twocolumn,showpacs,floatfix,english,noshowpacs]{revtex4}

\usepackage{url}
\usepackage{bm,bbm}
\usepackage{graphicx}
\usepackage{color}
\usepackage[colorlinks=true, urlcolor=blue, linkcolor=blue, citecolor=blue, pdftex]{hyperref}
\usepackage[english]{babel}

\usepackage{soul}
\usepackage{blindtext}

\newcommand{\dagga}{{\phantom{\dagger}}}

\begin{document}

\title{Variational wave functions for the spin-Peierls transition in the Su-Schrieffer-Heeger model with quantum phonons}

\author{Francesco Ferrari}
%\email[]{ferrari@itp.uni-frankfurt.de}
\affiliation{Institute for Theoretical Physics, Goethe University Frankfurt, Max-von-Laue-Stra{\ss}e 1, D-60438 Frankfurt a.M., Germany}
\author{Roser Valent\'i}
\affiliation{Institute for Theoretical Physics, Goethe University Frankfurt, Max-von-Laue-Stra{\ss}e 1, D-60438 Frankfurt a.M., Germany}
\author{Federico Becca}
\affiliation{Dipartimento di Fisica, Universit\`a di Trieste, Strada Costiera 11, I-34151 Trieste, Italy}

\date{\today}

\begin{abstract}
We introduce variational wave functions to evaluate the ground-state properties of spin-phonon coupled systems described by the Su-Schrieffer-Heeger
model. Quantum spins and phonons are treated on equal footing within a Monte Carlo sampling, and different regimes are investigated. We show that 
the proposed variational {\it Ansatz} yields good agreement with previous density-matrix renormalization group results in one dimension and is able 
to accurately describe the spin-Peierls transition. This variational approach is neither constrained by the magnetoelastic-coupling strength nor by 
the dimensionality of the systems considered, thus allowing future investigations in more general cases, which are relevant to spin-liquid and 
topological phases in two spatial dimensions.
\end{abstract}

\maketitle

\section{Introduction}

Effective models on the lattice constitute an invaluable tool to describe the low-energy properties of condensed-matter systems. Here, the 
original problem of interacting electrons and ions is simplified by keeping a few ``relevant'' (Wannier) orbitals on each atom, with a reduced 
number of effective couplings, the most notable one being the Hubbard-$U$~\cite{georges2013}. In most cases, the Born-Oppenheimer approximation 
is adopted, implying a static lattice structure. However, lattice vibrations (e.g., phonons) may play a fundamental role in determining the 
actual low-energy properties of the system. In this respect, superconductivity represents the most striking example~\cite{bardeen1957}, where 
the attractive interaction among electrons is mediated by phonons. In this regard, one can replace the retarded phonon-mediated interaction by 
an instantaneous attraction among electrons in a purely fermionic model, e.g., the negative-$U$ Hubbard model~\cite{nozieres1985}. Besides 
superconductivity, phonons may also induce other kind of electron instabilities, such as charge-density waves that are triggered by Peierls 
distortions of the underlying lattice~\cite{peierls1955}. In general, attacking the full problem of coupled electrons and phonons is not easy, 
even when this is limited to some effective model, e.g., Fr\"ohlich~\cite{frohlich1954} and Holstein~\cite{holstein1959} ones. With the advent 
of efficient numerical algorithms, however, there has been an increasing number of investigations of models that explicitly include phonon 
degrees of freedom in one and two spatial dimensions, also in connection to high-temperature 
superconductors~\cite{hohenadler2004,clay2005,assaad2008,nowadnick2012,hohenadler2013,ohgoe2014,ohgoe2017,karakuzu2017,nomura2020,costa2020}.

Besides the Fr\"ohlich and Holstein models, in which phonons are coupled to the electron density, the Su-Schrieffer-Heeger (SSH)~\cite{su1979} 
model represents an important alternative to study the effect of phonons on the electronic properties. The SSH model was introduced to describe the 
soliton formation in one-dimensional systems, like Polyacetylene~\cite{su1979}; here, lattice displacements are directly coupled to the electronic 
hopping. There are three possible ways to treat the lattice deformations within the SSH model, with increasing complexity. The simplest approach 
considers a static modulation of the hopping amplitudes, thus avoiding degrees of freedom for the lattice. This way to proceed has been widely 
explored in the recent past as a simple model for topological insulators~\cite{asboth2016}. Within an adiabatic approximation, where the kinetic
energy of phonons is neglected, lattice distortions can be treated as classical variables and optimized to find the best energy state (in the 
presence of an elastic energy for each site). Finally, the hardest approach is considering the full quantum dynamics of phonons. In the latter case, 
investigations have been limited to one-dimensional systems~\cite{fradkin1983,sengupta2003,weber2015}, even though applications to two-dimensional 
systems have been proposed~\cite{alder1997,li2020}. 

Phonons are also relevant in Mott insulators. Here, the super-exchange coupling $J$, like the electron hopping, depends on the distance between 
ions and, therefore, the spin-spin interaction is directly affected by phonons. In this case, a relevant low-energy model is the SSH model for 
spins, with the hopping operator replaced by the bilinear Heisenberg interaction. The inorganic compounds CuGeO$_3$~\cite{hase1993,muthukumar1997}, 
NaV$_2$O$_5$,~\cite{isobe1996,gros1999} and TiOCl~\cite{seidel2003,hoinkis2005} are typical examples where phonons drive a spin-Peierls 
transition~\cite{boucher1996,lemmens2003}. In the adiabatic approximation, the one-dimensional spin-$1/2$ Heisenberg model coupled to classical 
displacement variables is unstable with respect to a static dimerization, no matter how small the spin-phonon coupling is~\cite{cross1979}. This 
is because the energy gain for a distortion is linear in the displacement, while the loss due to the elastic energy is quadratic. Then, the 
distortion immediately leads to a spin gap in the excitation spectrum. The adiabatic limit has been studied in detail for a variety of 
cases~\cite{feiguin1997,augier1998,augier2000,becca2003,zhang2008}. Here, while the spin degrees of freedom retain their quantum character, 
lattice displacements are treated classically with a relatively small increase of the computational cost with respect to the pure spin model.

Going beyond this approximation and treating quantum phonons is not an easy task. The main reason is due to the explosion of the Hilbert space, 
since an arbitrarly large number of phonons may exist on each lattice site. This fact poses serious problems in numerical diagonalizations or 
density-matrix renormalization group approaches, where a truncation of the Hilbert space is required~\cite{wellein1998,bursill1999,pearson2010}. 
While this kind of approximation may be justified for large phonon frequencies $\omega$ (i.e., $\omega \gg J$), in the opposite limit 
$\omega \ll J$, several phonons may be necessary to have an accurate description of the ground state. Perturbation expansion and effective spin 
models may be also pursued~\cite{uhrig1998,weisse1999}, but again the generic case with $\omega \approx J$ cannot be assessed. Quantum Monte Carlo 
methods~\cite{sandvik1999} do not have limitations coming from the infinite Hilbert space of phonons, but they can be employed only in cases where 
the Hamiltonian has no sign problem, thus having a limited applicability. 

The need to consider the full quantum model comes from the fact that in most materials (e.g., CuGeO$_3$) the phonon frequency is of the same order 
of magnitude of $J$. Thus, away from the adiabatic limit $\omega/J=0$, the properties of the system may be largely affected by phonon dynamics. 
For example, a finite spin-phonon coupling is needed to drive the system into a gapped (dimerized) state. Indeed, since the phonon displacement is 
coupled to the dimerization operator, a small spin-phonon perturbation gives rise to a next-nearest-neighbor spin-spin interaction $J^\prime$ and 
it is well known that a finite $J^\prime/J$ is needed to open a spin gap in one dimension~\cite{uhrig1998,weisse1999,private}. These arguments have 
been confirmed by accurate density-matrix renormalization group calculations~\cite{bursill1999} and Monte Carlo simulations~\cite{sandvik1999}.
In addition, including vibrations and displacements of the lattice is important for several magnetic materials, either in magnetically ordered
phases (where phonons may affect the magnon dispersion) or in absence of magnetic long-range order (where phonons stand up in the competition 
between valence-bond solids and spin liquids). 

In this work, we devise variational wave functions, which can be treated within a Monte Carlo sampling, to assess the ground-state properties of
the spin SSH model in one dimension:
\begin{align}\label{eq:1dham_bosons}
 \mathcal{H}=J\sum_{i=1}^L  &\left[1+ g (a_{i+1}+a_{i+1}^\dagger-a_i-a_i^\dagger) \right] \mathbf{S}_i \cdot \mathbf{S}_{i+1} \nonumber \\
 &+\omega\sum_{i=1}^L \left(a_i^\dagger a_i^\dagga+\frac{1}{2}\right).
\end{align}
Here, ${\bf S}_{i}$ is the spin-$1/2$ operator, $a^\dag_{i}$ ($a_{i}$) is the creation (annihilation) phonon operator on the site $i$, and $L$ is 
the total number of lattice sites (periodic boundary conditions are considered). The physics of the system is governed by the values of the bare 
(antiferromagnetic) super-exchange constant $J$, the magnetoelastic coupling strength $g$ and the frequency of the Einstein phonons $\omega$.
We will show that a suitably defined variational approach is able to reproduce the ground-state properties of the system for different regimes, 
with the adiabatic parameter $\omega/J$ ranging from $0.1$ to $10$. Our results are found to be in very good agreement with previous density-matrix 
renormalization group calculations~\cite{bursill1999}. As we discuss in the conclusions, the present variational method can be extended to 
higher-dimensional spin-phonon problems.

The paper is organized as follows: In section~\ref{sec:method}, we describe the variational Monte Carlo method, in section~\ref{sec:results}, we
present the numerical results, and in section~\ref{sec:concl}, we draw our conclusions.

\section{The variational method}\label{sec:method}

Within our variational framework, we approximate the ground-state wave function of the Hamiltonian~\eqref{eq:1dham_bosons} by a correlated 
variational {\it Ansatz} that is the product of a spin wave function, $|\Psi_s\rangle$, a phonon wave function, $|\Psi_p\rangle$, and a spin-phonon
Jastrow factor, $\mathcal{J}_{sp}$, which couples the two different degrees of freedom of the system:
\begin{equation}\label{eq:psivar}
 |\Psi_0\rangle=\mathcal{J}_{sp} |\Psi_s\rangle \otimes |\Psi_p\rangle.
\end{equation}
Both spins and phonons are treated at the quantum level, and the expectation values of the physical observables are computed by performing a Monte 
Carlo sampling of the (infinitely large) Hilbert space. The configurations of the system which are visited by the Markov chain are labelled by the 
local spin and phonon states of each lattice site. For the local Hilbert space of the spins we adopt the conventional choice of labeling the states 
by the $S^z_j$ quantum number. Regarding the phonon degrees of freedom, instead, we will consider two alternative local quantum numbers, namely the 
number of phonons ${n_j=a^\dagger_j a_j}$ (discrete label) and the site displacement ${X_j= a_j + a^\dagger_j}$ (continuous variable).

The spin wave function $|\Psi_s\rangle$ entering the variational {\it Ansatz} of Eq.~\eqref{eq:psivar} is a Gutzwiller-projected fermionic state 
of the form:
\begin{equation}
 |\Psi_s\rangle= \mathcal{J}_{ss} \mathcal{P}_G |\Phi_0\rangle.
\end{equation}
Here, $|\Phi_0\rangle$ is the ground state wave function of an auxiliary BCS Hamiltonian of Abrikosov fermions, which contains hopping and singlet 
pairing terms. The application of the Gutzwiller projector, $\mathcal{P}_G$, to the fermionic state $|\Phi_0\rangle$ yields a suitable wave function
for spins. The parameters of the BCS Hamiltonian, i.e. the hopping and pairing amplitudes, play the role of variational parameters. More details 
concerning the fermionic wave functions can be found in Ref.~\cite{becca2009,ferrari2018}. In addition, a spin-spin Jastrow factor is included, 
\begin{equation}
 \mathcal{J}_{ss}=\exp\left[\sum_{i,j} v_{s}(i,j) S^z_i S^z_j\right],
\end{equation}
whose {\it pseudopotential} parameters $v_{s}(i,j)=v_{s}(|R_i-R_j|)$ depend only on the relative distance of the sites in the undistorted spin chain.

The uncorrelated phononic part of the variational wave function~\eqref{eq:psivar} is a coherent state for the phonon mode with momentum $k$:
\begin{equation}\label{eq:phonwf}
 |\Psi_p\rangle=\exp(z a_k^\dagger) |0\rangle_p =\prod_j \exp(z e^{i k R_j} a_j^\dagger) |0\rangle_p
\end{equation}
Here $|0\rangle_p$ is the vacuum state of phonons and $R_j$ is the (integer) equilibrium coordinate of site $j$. The real variable $z$ is a fugacity
variational parameter which determines the average number of phonons per site
\begin{equation}
 \langle n_j \rangle_p  =\frac{\langle \Psi_p| a_j^\dagger a_j |\Psi_p\rangle}{\langle \Psi_p|\Psi_p\rangle}=z^2,
\end{equation}
and the amplitude of the site displacements
\begin{equation}
 \langle X_j \rangle_p  =\frac{\langle \Psi_p| (a_j+a_j^\dagger) |\Psi_p\rangle}{\langle \Psi_p|\Psi_p\rangle}=2z\cos(k R_j).
\end{equation}
The momentum $k$ of the phonon mode modulates the direction of sites displacements. The Peierls instability of the spin SSH chain towards 
dimerization is achieved by taking $k=\pi$. 

Depending on how we choose to represent the local Hilbert space of phonons, we can have different spin-phonon Jastrow factors. On the one side, 
by using the computational basis labeled by local phonon numbers on each site (i.e., $n_j$), as done in Ref.~\cite{karakuzu2017}, we can take:
\begin{equation}\label{eq:jsp_ssn}
 \mathcal{J}_{sp}=\mathcal{J}_{n}=\exp\left(\sum_{i,j} v_{n}(i,j) S^z_i S^z_j n_j\right),
\end{equation}
where $v_{n}(i,j)=v_{n}(|R_i-R_j|)$ is another set of translationally invariant pseudopotential parameters. Within this choice, the uncorrelated
phonon wave function $|\Psi_p\rangle$ is rewritten as a linear superposition of the many-body configurations $|n_1,\dots,n_L\rangle$, which are 
then sampled by Monte Carlo:
\begin{equation}
 |\Psi_p\rangle=\sum_{n_1,\dots,n_L} \frac{z^{N_p} e^{ik\sum_j R_j n_j}}{\sqrt{n_1!\cdots n_L!}} |n_1,\dots,n_L\rangle.
\end{equation}
Here, $N_p=\sum_j n_j$ is the total number of phonons in the chain. As shown in the next section, it turns out that this variational state is not 
the optimal choice for the model under investigation and considerably better results are obtained by employing a Jastrow factor in which the spins 
are coupled to the relative displacements of the lattice sites. 

Indeed, we can adopt a different computational basis, which is diagonal in $X_j$, similarly to what is done in 
Ref.~\cite{ohgoe2014,ohgoe2017}, and introduce the following spin-phonon Jastrow factor:
\begin{equation}\label{eq:jsp_ssx}
 \mathcal{J}_{sp}=\mathcal{J}_{X}=\exp\left[\frac{1}{2}\sum_{i,j} v_{X}(i,j) S^z_i S^z_j(X_i-X_j)\right].
\end{equation}
In this case the pseudopotential parameters $v_{X}(i,j)=v_{X}(R_i-R_j)$ are still assumed to be translationally invariant, but they are odd with 
respect to the exchange of lattice sites. Accordingly, we can reformulate the uncorrelated phononic part $|\Psi_p\rangle$ in terms of the 
many-body configurations $|X_1,\dots,X_L\rangle$ as follows 
\begin{equation}
  |\Psi_p\rangle=\int dX_1 \cdots dX_L \left[ \prod_j e^{\phi_j(X_j)} \right] |X_1,\dots,X_L\rangle,
\end{equation}
where
\begin{equation}
 \phi_j(X_j) = iz\sin(kR_j)X_j-\frac{1}{4}[X_j-2z\cos(kR_j)]^2.
\end{equation}
We would like to stress the fact that, within this approach, a cutoff on the number of phonons is not required, in contrast to the case of
Ref.~\cite{ohgoe2014,ohgoe2017}, where a different uncorrelated phonon state is employed. The use of the spin-displacement Jastrow factor 
$\mathcal{J}_{X}$ provides a remarkable accuracy gain with respect to the Jastrow factor of Eq.~\eqref{eq:jsp_ssn}, see below. As a consequence 
of this change of paradigm, we adopt a suitable Monte Carlo scheme in which we sample the Hilbert space of the phonons by specifying the 
displacements of the lattice sites, $\{X_j\}$. For this reason, we conveniently rewrite the problem of Eq.~\eqref{eq:1dham_bosons} by replacing 
the bosonic creation and annihilation operators with the (adimensional) displacement and momentum operators, ${X_j=(a_j^\dagger+a_j)}$ and 
${P_j=i(a_j^\dagger-a_j)}$, which satisfy ${[X_j,P_j]=2i}$. The Hamiltonian takes the alternative form
\begin{align}\label{eq:1dham_coord}
 \mathcal{H}=J\sum_{i=1}^L  &\left[1+ g (X_{i+1}-X_i) \right] \mathbf{S}_i \cdot \mathbf{S}_{i+1} \nonumber \\
 &\qquad \quad +\frac{\omega}{4}\sum_{i=1}^L \left[P_i^{2} + X_i^{2}\right].
\end{align}
When computing the variational energy, the momentum operator acts as a derivative with respect to the displacement, namely 
$P_j=-2i \frac{\partial}{\partial X_j}$.

%%%%%%%%%%%%%%%%%%%%%%%%%%%%%%%%%%%%%%%%%%%%%%%%%%%%%%%%%%%%%%%%%%%%%%%%%%%%%%%%%%%%%%%%%%%%%%%%%%%%%%%%%%%%%%
\begin{figure}
\includegraphics[width=0.83\columnwidth]{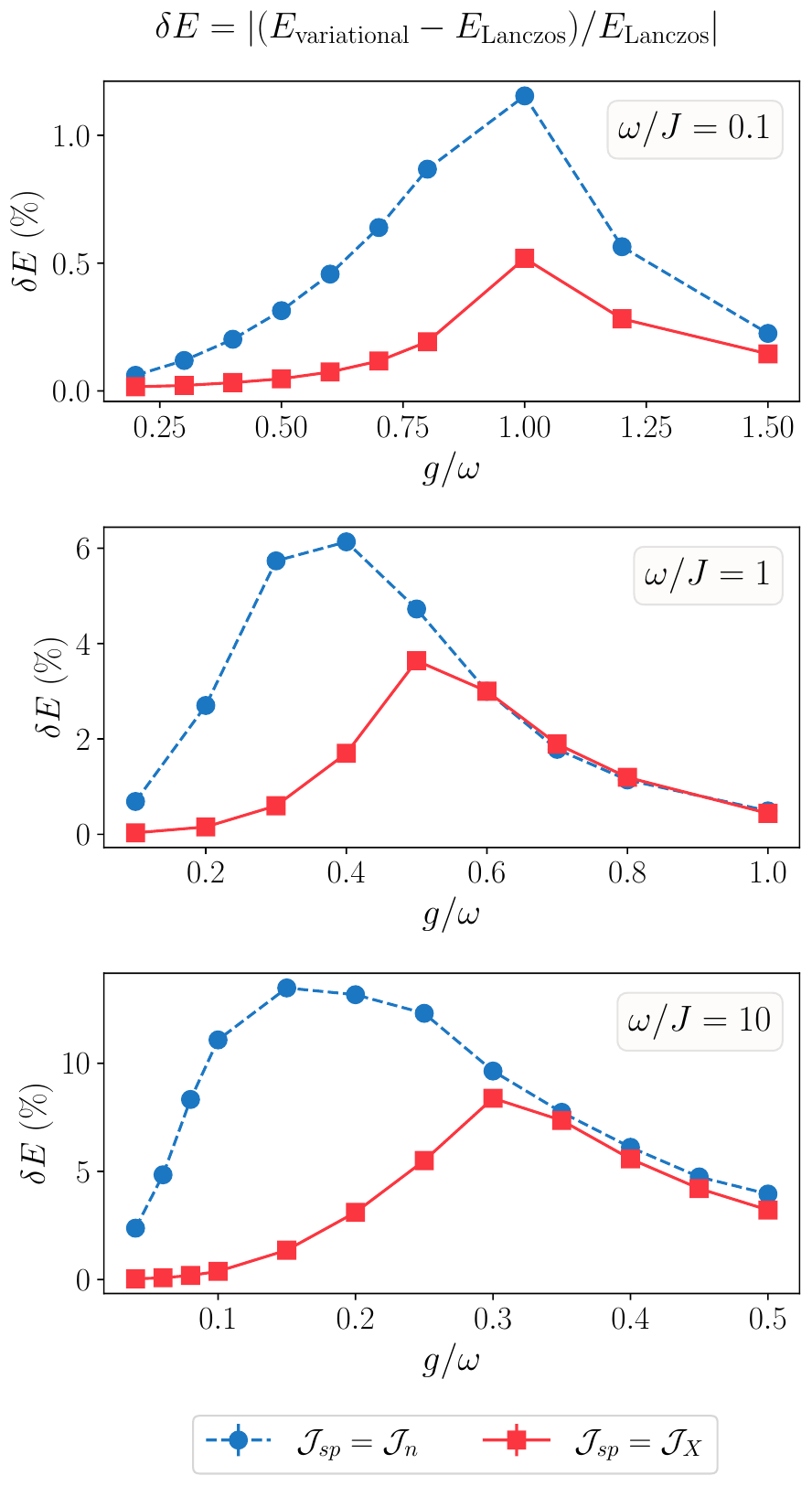}
\caption{\label{fig:accu} Relative error of the variational energies with respect to Lanczos results ($\delta E$) for a chain of $L=8$ sites.
The relative error is plotted (in percentage) as a function of $g/\omega$ for three values of the adiabatic parameter, $\omega/J=0.1$ (upper panel), 
$\omega/J=1$ (middle panel), and $\omega/J=10$ (lower panel). We note that the scale of the vertical axis is different in the three panels. 
Two sets of data are shown: blue circles represent the results obtained with the $\mathcal{J}_n$ spin-phonon Jastrow [Eq.~\eqref{eq:jsp_ssn}], 
while red squares correspond to the results obtained with the $\mathcal{J}_X$ spin-displacement Jastrow [Eq.~\eqref{eq:jsp_ssx}]. 
Error bars are smaller than the size of the dots.}
\end{figure}

\begin{figure*}
\includegraphics[width=2\columnwidth]{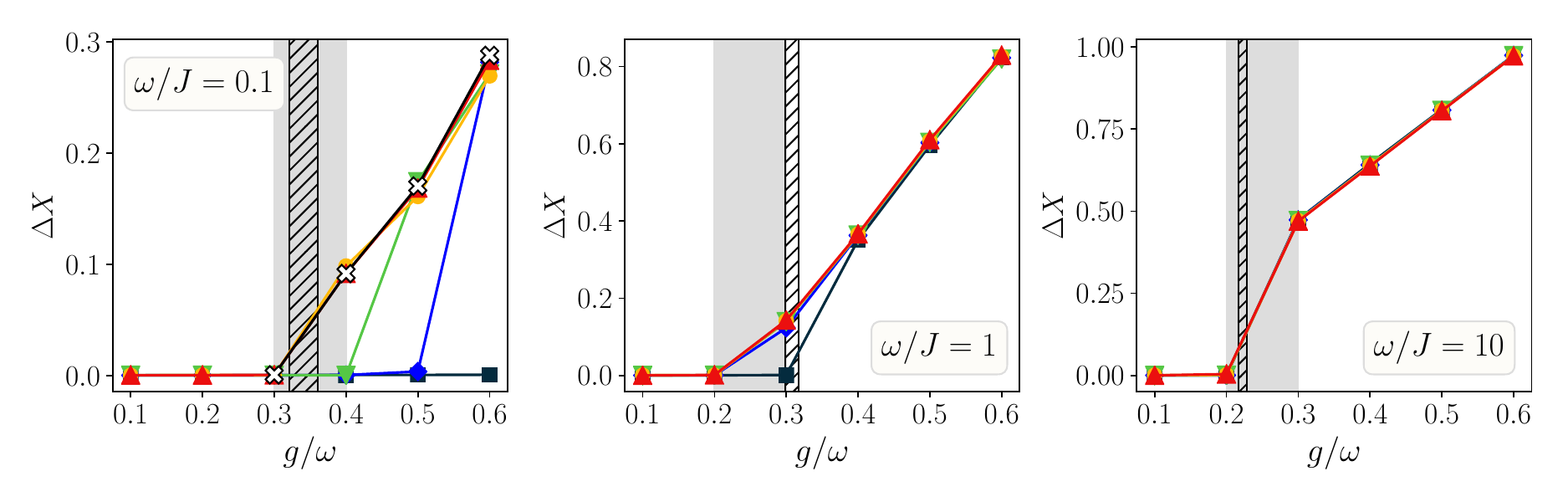}
\includegraphics[width=1.6\columnwidth]{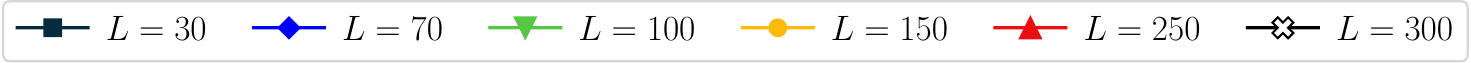}
\caption{\label{fig:deltax} Order parameter for the lattice deformation $\Delta X$ [Eq.~\eqref{eq:deltax}] as a function of $g/\omega$. 
Different lattice sizes $L$ are considered, as well as three different regimes, $\omega/J=0.1$ (left panel), $\omega/J=1$ (middle panel), 
and $\omega/J=10$ (right panel). Error bars are smaller than the size of the dots. The grey shaded area marks the region in which $\Delta X$
becomes finite in the thermodynamic limit. The hatched area denotes the position of the critical point (and its uncertainity) according to 
density-matrix renormalization group calculations~\cite{bursill1999}.}
\end{figure*}
%%%%%%%%%%%%%%%%%%%%%%%%%%%%%%%%%%%%%%%%%%%%%%%%%%%%%%%%%%%%%%%%%%%%%%%%%%%%%%%%%%%%%%%%%%%%%%%%%%%%%%%%%%%%%% 

\section{Results}\label{sec:results}

We apply our variational scheme to three different regimes of the spin SSH model, namely $\omega/J=0.1$ (adiabatic regime), $\omega/J=1$ and 
$\omega/J=10$ (anti-adiabatic regime). In order to correctly describe the spin-Peierls dimerization of the model, we consider a phonon coherent 
state~\eqref{eq:phonwf} with $k=\pi$. The optimal {\it Ansatz} for the variational wave function for the spins, $|\Psi_s\rangle$, is obtained 
by Gutzwiller-projecting the ground state of a BCS Hamiltonian with hopping and pairing terms at first- and second-neighboring 
sites~\cite{ferrari2018}. Since the phonon wave function~\eqref{eq:phonwf} breaks the translational invariance (for $z \ne 0$), we allow the 
first-neighbor couplings of the BCS Hamiltonian to take different values on the bonds $(2j,2j+1)$ and $(2j+1,2j+2)$, thus breaking the translations
also within the spin part of the wave function. This parametrization is suitable to describe the spin-Peierls phase, where spins form singlets on 
alternating bonds. All the parameters are numerically optimized by applying the stochastic reconfiguration technique~\cite{sorella2005}. 

%%%%%%%%%%%%%%%%%%%%%%%%%%%%%%%%%%%%%%%%%%%%%%%%%%%%%%%%%%%%%%%%%%%%%%%%%%%%%%%%%%%%%%%%%%%%%%%%%%%%%%%%%%%%%%
\begin{figure}
\includegraphics[width=\columnwidth]{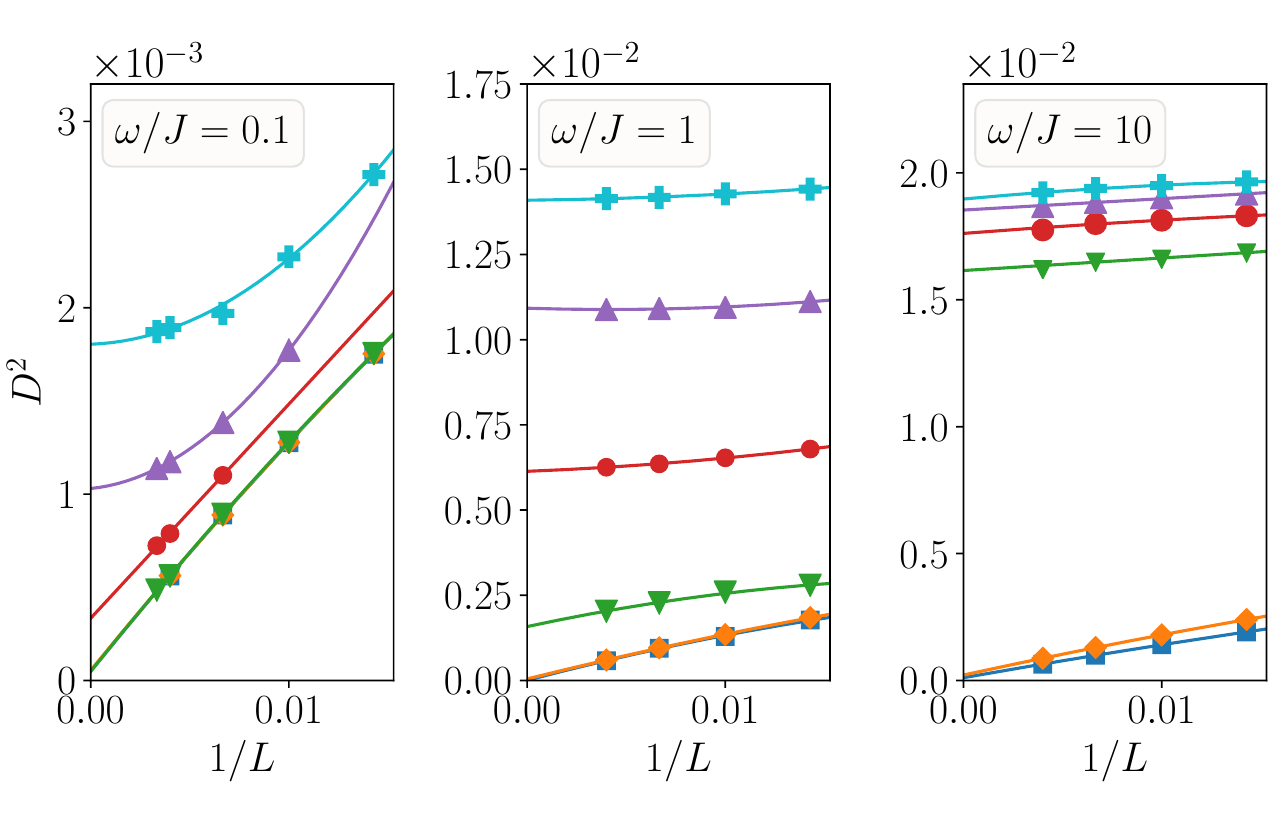}
\includegraphics[width=0.75\columnwidth]{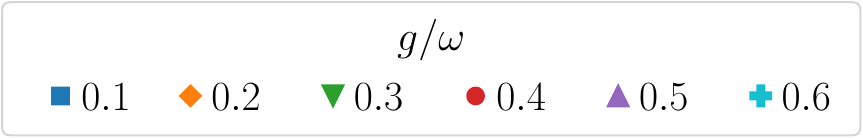}
\caption{\label{fig:dimer} Finite-size scaling of the Fourier-transformed dimer-dimer correlations $D^2$ [Eq.~\eqref{eq:dpi}]. Results for 
${\omega/J=0.1}$ (left panel), ${\omega/J=1}$ (middle panel), and ${\omega/J=10}$ (right panel) are shown. We note that the scale of the vertical 
axis is different in the first panel to account for the different order of magnitude of the correlation functions. The error bars are smaller than 
the size of the dots.}
\end{figure}
%%%%%%%%%%%%%%%%%%%%%%%%%%%%%%%%%%%%%%%%%%%%%%%%%%%%%%%%%%%%%%%%%%%%%%%%%%%%%%%%%%%%%%%%%%%%%%%%%%%%%%%%%%%%%%

We first assess the effectiveness of the $\mathcal{J}_n$ and $\mathcal{J}_X$ spin-phonon Jastrow factors, by comparing the variational energies 
with the ones obtained by Lanczos diagonalization on a finite cluster. Due to the infinitely large Hilbert space of phonons, the application of 
the Lanczos method requires a truncation of the Hilbert space. We adopt a truncation scheme in which we consider only phonon configurations 
$|n_1,...,n_L\rangle$ with ${n_j<n_{\rm max}}$ for each lattice site $j$. Within this approximation, we compute the Lanczos ground-state energies 
of the Hamiltonian~\eqref{eq:1dham_bosons} for a chain of $L=8$ sites. For this cluster, a threshold of $n_{max}=5$ ensures a satisfactory 
convergence of the energy in the range of parameters we considered. The results of the benchmark are summarized in Fig.~\ref{fig:accu}, where 
the relative error of the variational energy, $\delta E=\left|(E_{\rm variational}-E_{\rm Lanczos})/ E_{\rm Lanczos}\right|$, is plotted as a 
function of $g/\omega$ for the three cases $\omega/J=0.1$, $1$, and $10$. We first notice that accurate variational energies are achieved for 
small values of the adiabatic parameter $\omega/J$, where the dimerization due to the spin-Peierls instability is weaker (see below). Most 
importantly, we observe that the spin-phonon Jastrow factor $\mathcal{J}_X$, in which the $S^z$ spin operators are coupled to site displacements, 
provides considerably more accurate results than the case with $\mathcal{J}_n$. This is especially true for small and intermediate values of 
$g/\omega$, which are relevant to assess the phase transition between the gapless (not dimerized) and gapped (dimerized) phases. These results 
suggest that the variational {\it Ansatz} with the Jastrow factor $\mathcal{J}_X$ is the optimal choice when the spins (or electrons) are coupled 
to the phonons through the relative displacement of lattice sites, as in the SSH model. Instead, the Jastrow factor of Eq.~\eqref{eq:jsp_ssn}, 
or an analogous version of it in which $n_j$ is replaced by $X_j$, could be the most suitable variational guess for the study of a spin-phonon 
model in which the the spins are coupled to bond phonons~\cite{sandvik1999,raas2002,weisse2006}. 

After having evaluated the degree of accuracy of our method, we consider larger clusters (up to $L=300$) and we study the phase transition from 
the gapless to the dimerized phase using the optimal variational {\it Ans\"atze} with the Jastrow factor $\mathcal{J}_X$. In order to locate the 
transition point, we can check the behavior of two different observables as a function of $g/\omega$. On the one hand, we can measure the net 
lattice deformation due to sites displacements by computing the order parameter~\cite{barford2005}
\begin{equation}\label{eq:deltax}
 \Delta X=\left |\frac{1}{L}\sum_{j=1}^L e^{i\pi R_j} \langle X_j \rangle_0 \right|,
\end{equation}
where $\langle \cdots \rangle_0={\langle \Psi_0 | \cdots | \Psi_0 \rangle}/{\langle \Psi_0 | \Psi_0 \rangle}$. $\Delta X$ becomes finite in the 
Peierls phase, where alternating short and long nearest-neighbor bonds are formed. The values of $\Delta X$ for various lattice sizes are reported 
in Fig.~\ref{fig:deltax}, in the different regimes under investigation ($\omega/J=0.1$, $1$, and $10$). We also compare in Fig.~\ref{fig:deltax} 
our estimates of the critical points $g_c/\omega$ with the predictions of density-matrix renormalization group calculations~\cite{bursill1999}. 
We point out that the latter estimates are obtained by a different approach based on the detection of a singlet-triplet level crossing in the 
low-energy spectrum, which is more accurate than looking at the order parameter. The computation of singlet and triplet excitations within our
variational approach requires a full optimization of these states, including the spin-phonon Jastrow factor, which is beyond the scope of the 
present work. 

\begin{figure}
\includegraphics[width=\columnwidth]{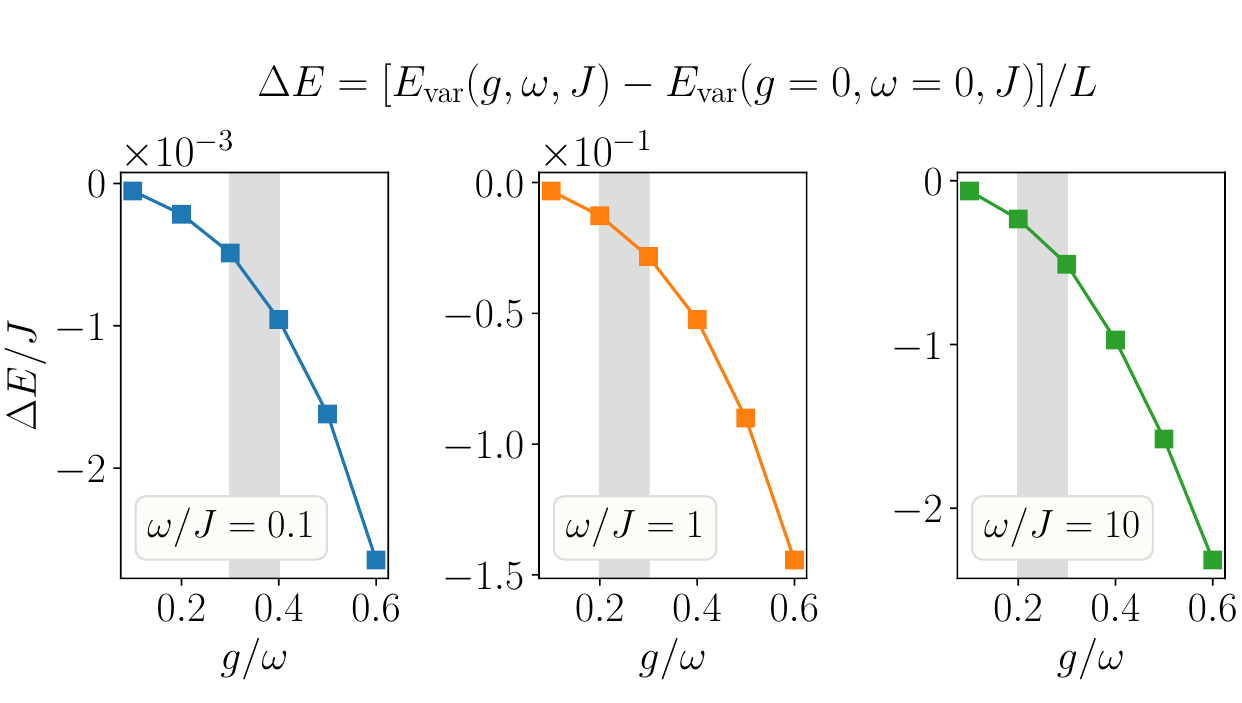}
\caption{\label{fig:enegain} Energy gain (in unit of $J$) due to the SSH coupling of the spins with phonons. The figure displays the difference 
between the variational energies of the full SSH system, $E_{\rm var}(g,\omega,J)$, and the one of the simple Heisenberg model, 
$E_{\rm var}(g=0,\omega=0,J)$, where no phonons are present. The results are obtained for a chain with $L=250$ sites as a function of $g/\omega$, 
for ${\omega/J=0.1}$ (left panel), ${\omega/J=1}$ (middle panel), and ${\omega/J=10}$ (right panel). We note that the scale of the vertical axis 
is different in the various panels in order to account for the different order of magnitude of the energy gain. As in Fig.~\ref{fig:deltax}, the 
grey shaded area marks the region in which we observe the onset of Peierls dimerization. The error bars are smaller than the size of the dots.}
\end{figure}

\begin{figure}
\includegraphics[width=\columnwidth]{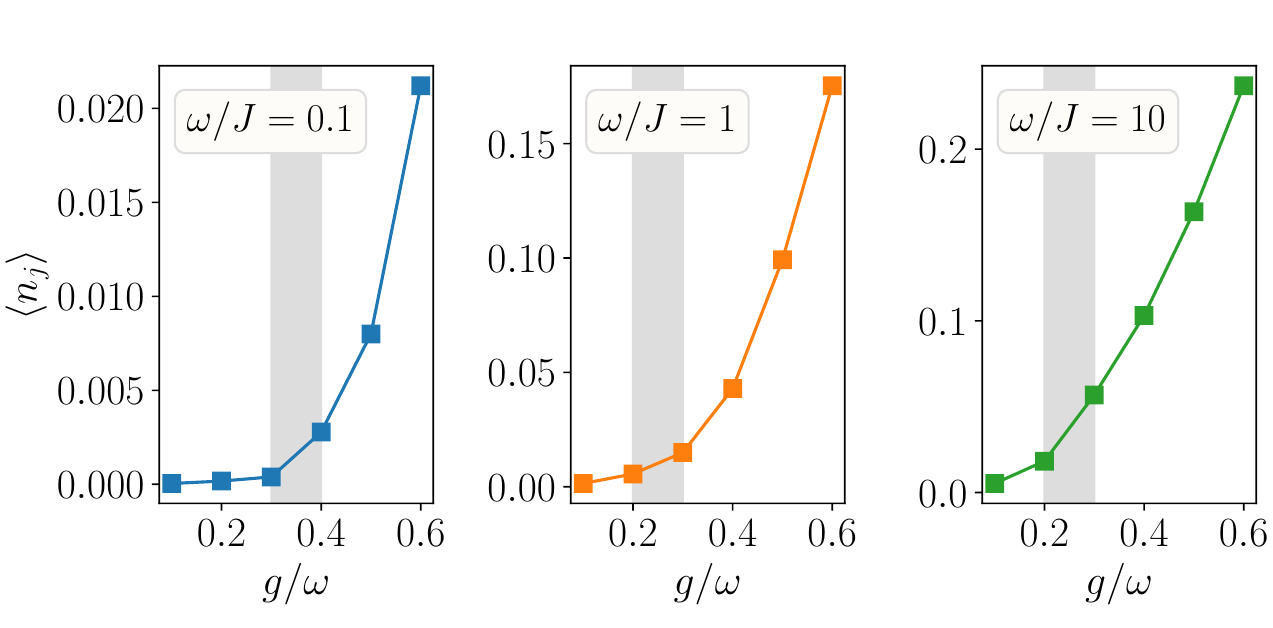}
\caption{\label{fig:phonnum} Average number of phonons per site [$\langle n_j \rangle$] as a function of $g/\omega$, for ${\omega/J=0.1}$ (left 
panel), ${\omega/J=1}$ (middle panel), and ${\omega/J=10}$ (right panel). The results refer to a chain with $L=250$ sites. As in 
Fig.~\ref{fig:deltax}, the grey shaded area marks the region in which we observe the onset of Peierls dimerization. The error bars are smaller 
than the size of the dots.}
\end{figure}

In addition to $\Delta X$, we also compute the Fourier-transformed dimer-dimer correlations at $k=\pi$ that help detect the presence of dimer order:
\begin{equation}\label{eq:dpi}
  D^2=\frac{1}{L}\sum_{R=0}^{L-1} e^{i\pi R}  \left(\frac{1}{L} \sum_{j=1}^L \langle S^z_j S^z_{j+1} S^z_{j+R} S^z_{j+R+1} \rangle_0 \right).
\end{equation}
A finite value of $D^2$ in the thermodynamic limit is a signal of spin dimerization. The finite size scaling analysis of this quantity is reported 
in Fig.~\ref{fig:dimer}. To further characterize the phase transition from the gapless to the dimerized phase, in Fig.~\ref{fig:enegain} we report 
the energy gain of the spin-phonon systems with respect to the Heisenberg limit (i.e., $g=0$ and $\omega=0$). Finally, in Fig.~\ref{fig:phonnum},
we show the average number of phonons per site. The order of magnitude of these quantities is substantially different in the three regimes of 
$\omega/J$ we considered.

We start our discussion of the results with the adiabatic regime, $\omega/J=0.1$. As shown in Fig.~\ref{fig:deltax}, the order parameter $\Delta X$ 
becomes finite in the interval ${0.3<g/\omega<0.4}$, in excellent agreement with the predictions of Ref.~\cite{bursill1999}. Within this regime, 
similarly to what is found in Ref.~\cite{bursill1999}, the results are strongly affected by finite size effects and large clusters ($L \gtrsim 150$ 
sites) are needed to reliably locate the phase transition. We note that, in general, for small values of $\omega/J$ the spin dimerization is very 
weak in the vicinity of the critical point. Indeed, both the dimer-dimer correlations (see Fig.~\ref{fig:dimer}) and the energy gain of the spin 
system due to the SSH coupling with phonons are relatively small, as shown in Fig.~\ref{fig:enegain}. A small number of phonons is involved
in the process of dimerization (see Fig.~\ref{fig:phonnum}). The situation is considerably different in the anti-adiabatic regime, $\omega/J=10$, 
where we observe a rapid increase of the dimer-dimer correlations $D^2$ just after the phase transition, which is the consequence of a strong 
dimerization of the spins. Here, the energy gain of the spin system due to the spin-phonon coupling is much larger than the one at $\omega/J=0.1$.
We locate the Peierls transition in the interval $0.2<g/\omega<0.3$, again in quantitative agreement with density-matrix renormalization group 
calculations~\cite{bursill1999}. This result is encouraging since the accuracy of the variational wave function is much deteriorated in comparison 
to the adiabatic limit, see Fig.~\ref{fig:accu}. We remark that, in contrast to the $J_1-J_2$ model (without phonons), where the continuous 
transition between gapless and gapped states can be described by using a fully-symmetric wave function~\cite{ferrari2018}, here the variational 
state explicity breaks the translational symmetry (see above), thus leading to less accurate description on any finite size; nevertheless, it is 
still possible to locate the phase transition with a good degree of precision. Finally, in the intermediate case with $\omega/J=1$, the onset of 
the spin dimerization appears for $0.2<g/\omega<0.3$, similarly to the case with $\omega/J=10$. This estimate, which is extracted both from the 
behavior of $\Delta X$ and the finite-size scaling analysis of the dimer-dimer correlations, is slightly different from the one obtained in 
Ref.~\cite{bursill1999}, which pinpointed the transition at $g/\omega \approx 0.31$. This discrepancy could be ascribed to the difficulies of 
reaching a sufficient accuracy in the intermediate regime with $\omega/J \approx 1$, i.e., when both spins and phonons have similar energy scales. 
We also mention that in the close proximity of the critical point two different variational {\it Ans\"atze}, a spin-fluid and a dimerized state, 
are extremely close in energy, and determining the optimal solution requires a very precise optimization of the variational parameters.

From these results, we conclude that, even though the best way to locate the transition between gapless and gapped phases is by looking at the 
singlet-triplet crossing, as done in Ref.~\cite{bursill1999}, a relatively accurate location of the Peierls transition, especially in the adiabatic
and anti-adiabatic regimes, may be also obtained from the analysis of the dimer-dimer correlations or the phonon displacement. We note, however, 
that these kind of calculations suffer from considerable size effects close to the phase transition. In particular, an accurate determination of 
the phonon displacements is remarkably hard for large sizes, so that a quantitative scaling analysis close to the Peierls instability is not 
possible. Nevertheless, the scope of this work was to demonstrate that a relatively simple variational wave function can capture the relevant 
features of the phase diagram of the SSH model.

%%%%%%%%%%%%%%%%%%%%%%%%%%%%%%%%%%%%%%%%%%%%%%%%%%%%%%%%%%%%%%%%%%%%%%%%%%%%%%%%%%%%%%%%%%%%%%%%%%%%%%%%%%%%%%

%%%%%%%%%%%%%%%%%%%%%%%%%%%%%%%%%%%%%%%%%%%%%%%%%%%%%%%%%%%%%%%%%%%%%%%%%%%%%%%%%%%%%%%%%%%%%%%%%%%%%%%%%%%%%%

\section{Conclusions}\label{sec:concl}

In this work, we analysed the spin-Peierls transition in the one-dimensional SSH model, where $S=1/2$ spins are coupled to quantum phonons, 
by using variational wave functions and Monte Carlo methods. In particular, we considered two ways to include the spin-phonon correlation through
Jastrow terms. The first one, which couples the spins to the phonon number, does not give accurate results, especially close to the spin-Peierls
transition. The second one, in which the Jastrow factor couples the spins to the sites displacements, provides a much better variational state.
Remarkably, in both cases no truncation in the Hilbert space of phonons is required. Our results show that this approach is able to describe
the phase transition between the gapless phase (for small values of the spin-phonon couplings) and the gapped one (for large values of $g/\omega$),
well reproducing previous density-matrix renormalization results~\cite{bursill1999}. Indeed, the agreement is excellent for both adiabatic (e.g.,
$\omega/J=0.1$) and anti-adiabatic (e.g., $\omega/J=10$) cases, while some minor discrepancies are obtained in the intermediate regime (e.g.,
$\omega/J=1$). Besides providing reliable calculations on the one-dimensional SSH model, our work paves the way for future investigations of
two-dimensional spin-phonon models, for which only very few accurate techniques are available at present. Indeed, although with a slightly reduced 
accuracy with respect to the one-dimensional case, Jastrow-fermionic wave functions can describe both magnetically ordered and disordered phases 
in higher-dimensional spin models~\cite{becca2009}, including spin-liquid phases and valence-bond solids in frustrated magnets. In particular,
the variational technique presented here can be employed to study the instability of spin-liquid phases towards the formation of valence-bond 
order as a consequence of the magnetoelastic coupling, which may play a relevant role in actual materials such as the spin-1/2 frustrated kagome 
compound herbertsmithite~\cite{norman2020,yingli2020,mendels2020} or the triangular lattice organic material 
$\kappa$-(ET)$_2$Cu$_2$(CN)$_3$~\cite{kanoda2017,riedl2019}.

\section*{Acknowledgments}

We thank F.F. Assaad, M. Fabrizio, C. Gros, N. Heinsdorf, and S. Sorella for useful discussion. F.F. acknowledges support from the Alexander von 
Humboldt Foundation through a postdoctoral Humboldt fellowship. R.V. acknowledges the Deutsche Forschungsgemeinschaft (DFG, German Research Foundation) for funding through TRR 288 - 
422213477  (project A05).

\end{document}